\documentclass[aps,prl,twocolumn,amsmath,showpacs,superscriptaddress]{revtex4}
\usepackage{graphicx,epsfig,longtable}
\usepackage{epstopdf}
\usepackage{dcolumn}
\usepackage{bm}

\def\gsimeq{\,\,\raise0.14em\hbox{$>$}\kern-0.76em\lower0.28em\hbox
{$\sim$}\,\,}
\def\lsimeq{\,\,\raise0.14em\hbox{$<$}\kern-0.76em\lower0.28em\hbox
{$\sim$}\,\,}
\def\be{\begin{equation}} 
\def\ee{\end{equation}}
\def\beqy{\begin{eqnarray}}
\def\eeqy{\end{eqnarray}}
\def\bmlet{\begin{mathletters}}
\def\emlet{\end{mathletters}}

\def\A&A#1#2#3{ \emph{ Astron. Astrophys. },  \textbf{#2}, #3  (#1). }

\begin{document}

\title{Skyrme-Hartree-Fock-Bogoliubov nuclear mass formulas:  Crossing the 0.6~MeV threshold with microscopically deduced pairing} 

\author{S.~Goriely}
\author{N.~Chamel}
\affiliation{Institut d'Astronomie et d'Astrophysique, CP-226, Universit\'e Libre de Bruxelles, 
1050 Brussels, Belgium}
\author{J.M.~Pearson}
\affiliation{D\'ept. de Physique, Universit\'e de Montr\'eal, Montr\'eal (Qu\'ebec), H3C 3J7 Canada}
\date{\today}

\begin{abstract}
We present a new Skyrme-Hartree-Fock-Bogoliubov nuclear-mass model in which the contact 
pairing force is constructed from microscopic pairing gaps of 
symmetric nuclear matter and neutron matter calculated from realistic 
two- and three-body forces, with medium-polarization effects included. With 
the pairing being treated more realistically than in any of our earlier
models, the rms deviation with respect to essentially all the available mass 
data falls to 0.581 MeV, the best value ever found within the mean-field 
framework. Since our Skyrme force is also constrained by the properties
of pure neutron matter this new model is particularly well-suited for 
application to astrophysical problems involving a neutron-rich environment,
such as the elucidation of the r-process of nucleosynthesis, and the 
description of supernova cores and neutron-star crusts.

\end{abstract}

\pacs{21.10.Dr,21.30.-x,21.60.Jz}

\maketitle

With a view to their astrophysical application in neutron-rich environments, 
we have developed a series of nuclear-mass models based on the Hartree-Fock-Bogoliubov 
(HFB) method with Skyrme and contact-pairing forces, together with phenomenological 
Wigner terms and correction terms for the spurious collective energy. 
All the model parameters are fitted to essentially all the experimental mass 
data.

Model HFB-9 \cite{sg05} and all later models constrained the underlying Skyrme
force to fit the energy-density curve of neutron matter, as calculated by 
Friedman and Pandharipande \cite{fp81} for realistic two- and three-nucleon
forces. In the latest of our published models, HFB-16 \cite{cha08}, we
imposed a comparable constraint on the contact pairing force. 
Instead of postulating a simple functional form for its density dependence, 
as is usually done, we constructed the pairing force by solving the HFB 
equations in uniform matter and requiring that the resulting gap reproduce 
exactly, as a function of density, the microscopic $^1S_0$ pairing gap 
calculated with realistic forces. In that preliminary study we assumed that
the pairing strength for neutrons (protons) depended only on the neutron 
(proton) density, as suggested by Duguet~\cite{dg04}, and chose for this microscopic 
reference gap  the one calculated for pure neutron matter without medium effects~\cite{lom01}.
We obtained thereby what was at the time our best-ever fit to the mass data, 
the rms deviation for our usual data set of 2149 measured masses of
nuclei with $N$ and $Z \ge$ 8 \cite{audi03} being 0.632 MeV. On the other hand,
the mass fits were much worse if we chose reference pairing gaps 
calculated with medium effects taken into account.  

Here we show that it is possible to obtain excellent mass fits even when the 
pairing force is constrained to microscopically calculated gaps in which 
medium effects have been included. The essential step is to impose the additional 
constraint of asymmetric nuclear matter pairing, thereby 
allowing the neutron and proton pairing strengths each to depend on both the 
neutron and proton densities.

{\it The HFB-17 mass model.} With this generalization of our earlier pairing
model,  Eq. (3.3) of Ref.~\cite{cha08} for the pairing strength is replaced by
\beqy
\label{eq.vpi}
& &v^{\pi\,q}[\rho_n, \rho_p]= -8\pi^2\left(\frac{\hbar^2}
{2 M_q^*(\rho_n, \rho_p)}\right)^{3/2} \times \nonumber \\ 
& &\qquad
\left(\int_0^{\mu_q+\varepsilon_{\Lambda}}{\rm d}\xi
\frac{\sqrt{\xi}}{\sqrt{(\xi-\mu_q)^2+\Delta_q(\rho_n, \rho_p)^2}}
\right)^{-1}  \quad ,
\eeqy
where $\Delta_q(\rho_n,\rho_p)$ is the corresponding pairing gap of
asymmetric nuclear matter calculated microscopically, 
$M_q^*(\rho_n, \rho_p)$ is the effective nucleon mass and 
$\varepsilon_{\Lambda}$ is the pairing cutoff. The chemical 
potential $\mu_q$ is approximated by $\mu_q = \hbar^2 k_{{\rm F}q}^2/(2 M_q^*)$,
where $k_{{\rm F}q} = (3\pi^2 \rho_q)^{1/3} $ is the Fermi wave number.

For the reference microscopic gap we use the recent Brueckner calculations of 
Cao et al.~\cite{cao06}, which are based on realistic two- and three-nucleon  
forces. All these calculations 
include the effect on the interaction of medium polarization, and
are performed both with and without self-energy corrections.
Ideally, we should have used the former gaps, which imply an effective mass
$M_q^*$ different from $M$. However, for complete consistency in 
Eq. (\ref{eq.vpi}) our Skyrme force would then have been required to
reproduce the same $M_q^*$. Satisfying this further constraint in addition
to all the others that we have already imposed seems to be impossible within 
the framework of the conventional Skyrme forces used here. Thus if one wishes
to retain the self-energy corrections the best that one can do is to relax
this constraint on the effective mass, but we found that this option was
incompatible with good mass fits.
We thus take the gaps calculated without self-energy
effects, and then for consistency set $M_q^* = M$ in Eq. (\ref{eq.vpi}). This 
was the choice that led to the excellent mass fits reported below.

\begin{figure}
\includegraphics[scale=0.3]{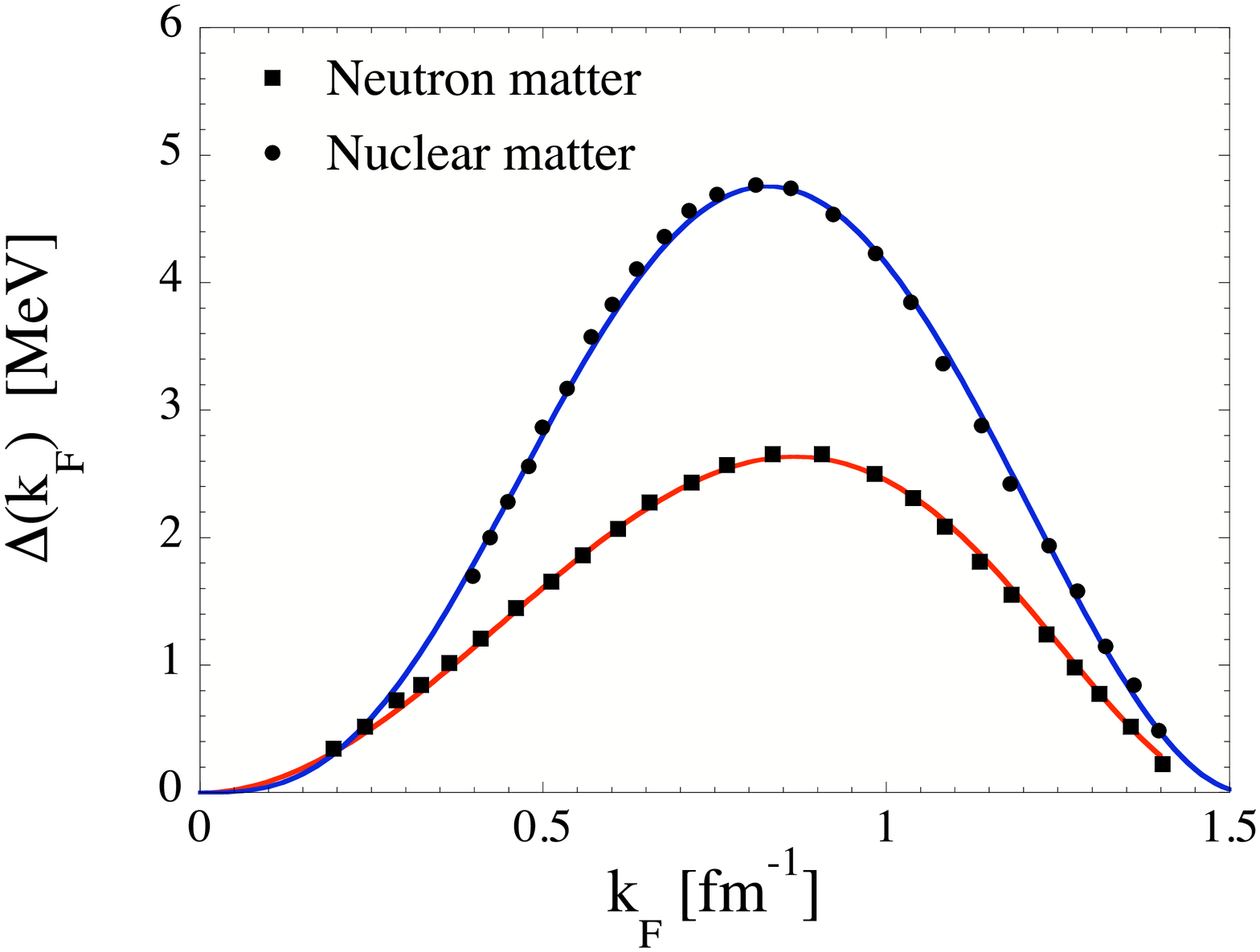}
\vskip -0.5cm
\caption{(Color online) $^1S_0$ pairing gap $\Delta$ in infinite neutron (square) and 
symmetric (circle) nuclear matter as a function of the Fermi wave number,
taken from Ref.~\cite{cao06}.}
\label{fig01}
\end{figure}

Ref.~\cite{cao06} calculates pairing gaps only for symmetric nuclear matter, 
$\Delta_{SM}(\rho = \rho_n+\rho_p)$, and pure neutron matter, 
$\Delta_{NM}(\rho_n)$. Since we need the pairing gaps for arbitrary asymmetry
we adopted the interpolation ansatz
\begin{eqnarray}
\Delta_q(\rho_n,\rho_p)=\Delta_{SM}(\rho)(1-|\eta|) \pm \Delta_{NM}(\rho_q)
\,\eta\,\frac{\rho_q}{\rho}   \, ,
\end{eqnarray}
where $\eta = (\rho_n-\rho_p)/\rho$ and the upper (lower) sign is to be taken 
for $q = n (p)$; we have also assumed charge symmetry, i.e., 
$\Delta_n(\rho_n,\rho_p) = \Delta_p(\rho_p,\rho_n)$. This expression ensures 
that for symmetric nuclear matter, $\Delta_q(\rho/2,\rho/2)=\Delta_{SM}(\rho)$ 
and for neutron matter $\Delta_n(\rho, 0)=\Delta_{NM}(\rho)$ 
and $\Delta_p(\rho, 0)=0$.

Because of Coulomb effects, and a possible charge-symmetry breaking of nuclear 
forces, we must allow for the proton pairing strength to be different from the
neutron pairing strength. Likewise, we follow our usual practice of allowing 
the pairing to be slightly stronger for nucleons of which there are an odd 
number. This procedure can be understood microscopically~\cite{ber08} as 
compensating for the neglect of the time-odd fields implicit in our use of the 
equal-filling approximation (EFA)~\cite{pmr08}. (Note that the odd nucleon will
nevertheless contribute to the time-even fields.)
We take account of these extra degrees of freedom by
multiplying the value of $v^{\pi\,q}[\rho_q]$, as determined through
Eq.~(\ref{eq.vpi}), by renormalizing factors $f^{\pm}_q$, where $f^+_p, f^-_p$
and $f^-_n$ are free, density-independent parameters to be included in the mass
fit.  We set $f^+_n = 1$, tacitly supposing that all effects related to charge-symmetry 
breaking act only on protons.

{\it Results.} The foregoing model, labeled HFB-17, was fitted to the above 
data set of 2149 
measured nuclear masses~\cite{audi03}; in making this fit we followed our
recently adopted strategy~\cite{gp08} of dropping the Coulomb-exchange term. 
The resulting parameter set, labeled BSk17, is given in Table~\ref{tab2} 
(definitions of these parameters can be found in \cite{cha08}).

\begin{figure}
\includegraphics[scale=0.3]{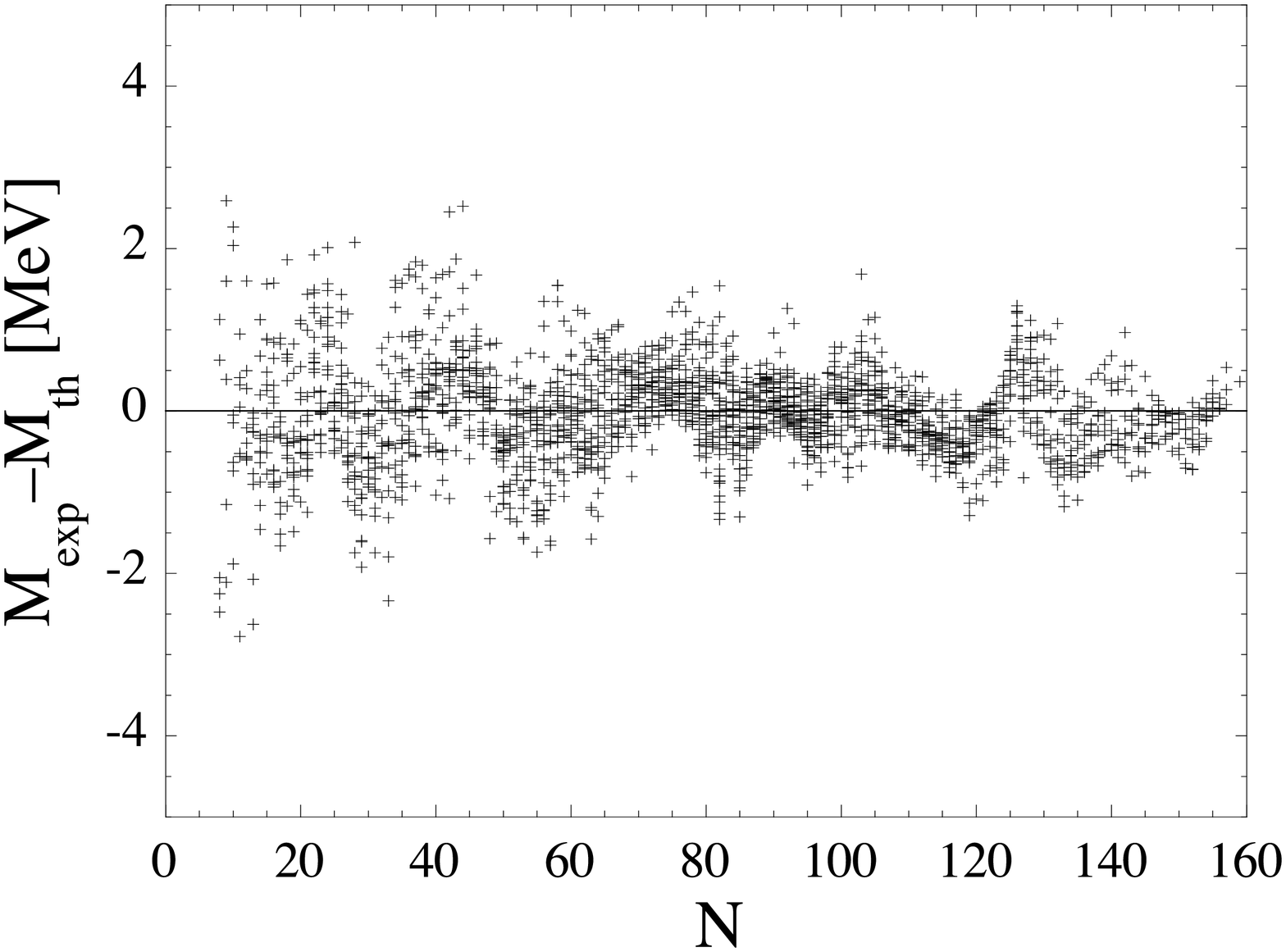}
\vskip -0.5cm
\caption{Differences between measured \cite{audi03} and HFB-17 masses, as 
function of $N$.}
\label{fig_mexp}
\end{figure}

The deviations (data-theory) between all the 2149 measured masses of our data 
set and the new predictions are shown graphically in Fig.~\ref{fig_mexp}; no
deviation exceeded 2.8 MeV.
The rms and mean values of these deviations are shown in the first two lines of
Table~\ref{tab4}; with an rms deviation of 0.581~MeV this is the most accurate 
mass model ever achieved within the mean-field framework. The next six lines
of this table refer to various subsets of our data set. 
We stress that all the 2149  data points to which we make our fit are taken 
from the 2003 Atomic Mass Evaluation (AME)~\cite{audi03}. However, a 
considerable amount of mass data has accumulated in the meantime, but since 
these new measurements have not been subjected to the same scrutiny that
went into the 2003 AME we have excluded them from our fit. Nevertheless, it is
of interest to compare these new data with our model, and we do this in lines
9 to 12 of Table~\ref{tab4} for two sets of measurements, Refs.~\cite{gsi05} 
and~\cite{jyf08}. It is remarkable that our model agrees better with these new 
data than with the fitting data.

\begin{table}
\centering
\caption{Parameter set BSk17: left panel shows the Skyrme parameters, middle panel
the pairing parameters and right panel the parameters for the Wigner and collective 
corrections (units for energy and length are MeV and fm respectively).} 
\label{tab2}
\vspace{.5cm}
\begin{tabular}{|l|c|}
\hline
  $t_0$  & -1837.33  \\    
  $t_1$  & 389.102 \\      
  $t_2$  & -3.1742 \\      
  $t_3$  & 11523.8  \\     
  $x_0$  &  0.411377  \\   
  $x_1$  & -0.832102 \\    
  $x_2$  & 49.4875  \\     
  $x_3$  &  0.654962  \\   
  $W_0$  &  145.885    \\  
  $\gamma$ &  0.3   \\
 \hline
\end{tabular}
\hskip 0.5cm
\begin{tabular}{|l|c|}
\hline
$f_{n}^+$  &  1.00  \\
$f_{n}^-$  &  1.04  \\
$f_{p}^+$  &  1.05  \\
$f_{p}^-$  & 1.05   \\
$\varepsilon_{\Lambda}$   &  16.0    \\
\hline
\end{tabular}  
\hskip 0.5cm
\begin{tabular}{|l|c|}
\hline
$V_W$ & -2.00   \\
$\lambda$                           & 320    \\
$V_W^{\prime}$  & 0.86   \\
$A_0$                               & 28    \\
$b$  &  0.8\\
$c$ & 10 \\
$d$ & 3.0 \\
$l$ & 14 \\
$\beta_2^0$ & 0.1\\
\hline
\end{tabular}       
\end{table}

We have pointed out in Ref. \cite{sg06} that only slight changes in the quality of the
mass fit can be accompanied by considerable changes in the overall pairing strength,
as measured by the spectral pairing gaps (defined in Eq. (5) of  Ref. \cite{sg06}). 
Now throughout our entire mass-model project we have been concerned not only with 
nuclear masses but also with nuclear level densities (among other quantities of
astrophysical interest), and these are very sensitive to the spectral pairing gaps.
The values of these quantities found here are comparable to those of model
HFB-13 \cite{sg06}, which lead to satisfactory level densities.

We have determined the ground-state spins and parities of odd nuclei  
using the HFB-17 single-particle level scheme obtained in the EFA, as described
above. For odd-A nuclei, the spin and parity are assumed to be those of the single nucleon of the last filled orbit, while for odd-odd nuclei we apply the Nordheim rule \cite{nor50}.
For the 90 spherical odd-A nuclei with quadrupole deformation parameter 
$\beta_2 \le 0.05$,   91\% of the experimental spins \cite{ripl2} are correctly predicted, while for 
the 717 deformed ones with $\beta_2 > 0.16$, only 41\% are correctly 
determined. For all the 1582 odd-A and odd-odd nuclei we obtain a global 
success rate of 47\% on the spins and 72\% on the parities 
(we assume a spherical configuration for $\beta_2 \le 0.16$).
The most important ground state properties predicted by the HFB-17
model, including spins and parities, have been
tabulated for all the 8508 nuclei with $8 \le Z\le 110$ between the proton and
neutron drip lines.
\begin{table}
\centering
\caption{Rms ($\sigma$) and mean ($\bar{\epsilon}$) deviations 
between data \cite{audi03} and HFB-17 predictions. The first pair of lines 
refers to all the 2149 measured masses $M$, the second pair to the masses 
$M_{nr}$ of the subset of 185 neutron-rich nuclei with $S_n \le $ 5.0 MeV, the 
third pair to the 1988 measured neutron separation energies $S_n$ and the
fourth pair to 1868 measured beta-decay energies $Q_\beta$. The fifth and six 
pairs correspond to the deviation with respect to the recently measured masses 
of Ref.~\cite{gsi05} and \cite{jyf08}, respectively. The seventh pair 
shows the comparison with the 782 measured charge radii \cite{ang04}, and the 
last line shows the calculated neutron-skin thickness of $^{208}$Pb.
Note that units for energy and length are MeV and fm respectively.}
\label{tab4}
\vspace{.5cm}
\begin{tabular}{|c|cc|}
\hline
&HFB-16&HFB-17 \\
\hline
$\sigma(2149~M)$  \cite{audi03}&0.632 &0.581   \\
$\bar{\epsilon}(2149~M)$  \cite{audi03}&-0.001 &-0.019 \\
$\sigma(M_{nr})$ \cite{audi03}&0.748& 0.729  \\
$\bar{\epsilon}(M_{nr})$  \cite{audi03}&0.161& 0.119 \\
$\sigma(S_n)$ \cite{audi03}&0.500&0.506 \\
$\bar{\epsilon}(S_n)$ \cite{audi03}&-0.012&-0.010\\
$\sigma(Q_\beta)$ \cite{audi03}&0.559&0.583 \\
$\bar{\epsilon}(Q_\beta)$ \cite{audi03}&0.031&0.022 \\
$\sigma(434~M)$  \cite{gsi05}&0.484&0.363   \\
$\bar{\epsilon}(434~M)$  \cite{gsi05}&-0.136&-0.092 \\
$\sigma(142~M)$  \cite{jyf08}&0.516 &0.548   \\
$\bar{\epsilon}(142~M)$  \cite{jyf08}&-0.070&0.172 \\
$\sigma(R_c)$ \cite{ang04}&0.0313&0.0300 \\
$\bar{\epsilon}(R_c)$ \cite{ang04}&-0.0149&-0.0114 \\
$\theta$($^{208}$Pb) &0.15&0.15\\
\hline
\end{tabular}
\end{table}

\begin{table}
\centering
\caption{Macroscopic parameters for forces BSk16 and BSk17. The first twelve 
lines refer to infinite nuclear matter, the last two to semi-infinite nuclear 
matter. Note that units for energy and length are MeV and fm respectively.}
\label{tab5}
\vspace{.5cm}
\begin{tabular}{|c|cc|}
\hline
& BSk16 & BSk17 \\
\hline
$a_v$ &-16.053&-16.054 \\
$\rho_0$ &0.1586&0.1586 \\
$J$ &30.0&30.0    \\
$M^*_s/M$ &0.80&0.80   \\
$M^*_v/M$ & 0.78&0.78      \\
$K_v$ &241.6&241.7\\
$L$ &34.87&36.28  \\
$G_0$ &-0.65&-0.69 \\
$G_0^{'}$&0.51&0.50  \\
$G_1$&1.52&1.55 \\
$G_1^{'}$ &0.44&0.45  \\
$\rho_{frmg}/\rho_0$&1.24&1.24\\
$a_{sf}$ &17.8&17.9  \\
$Q$ &39.0&38.1 \\
\hline
\end{tabular}
\end{table}

Table~\ref{tab5} shows the macroscopic parameters (infinite and semi-infinite 
nuclear matter) calculated for the force BSk17 (for the definition of these 
parameters see, for example, Ref.~\cite{cha08}). This table also shows the 
values of these parameters for force BSk16, the force underlying mass model
HFB-16, and it will be seen that in this respect there is very little 
difference between the two forces. This is hardly surprising given that the
macroscopic parameters depend entirely on the Skyrme force, and it is in the 
pairing channel that we have introduced the principal modifications. (Note
that the values of the isoscalar effective mass $M_s^*$ and symmetry energy $J$ 
were imposed.) It will be seen that in both models the isovector effective 
mass $M^*_v $ is found to be smaller than $M^*_s$ at the saturation density 
$\rho_0$, implying thereby that the neutron effective mass $M^*_n$ is larger 
than the proton effective mass $M^*_p$ in neutron-rich matter. Such an 
isovector splitting of the effective mass is consistent with measurements of 
isovector giant resonances~\cite{les06}, and has been confirmed in several 
many-body calculations with realistic forces~\cite{van05,zuo06}. 
 
\begin{figure}
\includegraphics[scale=0.3]{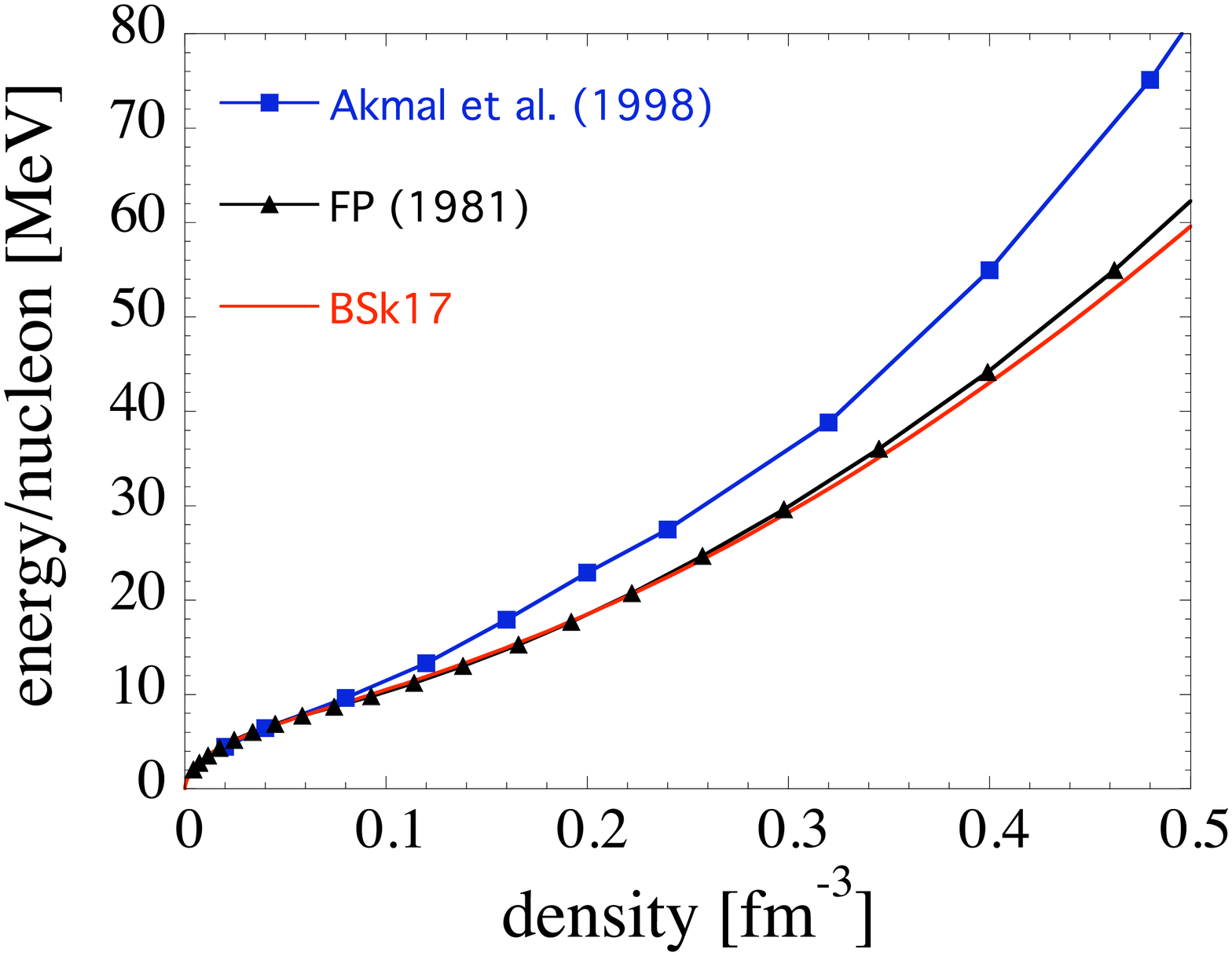}
\vskip -0.5cm
\caption{(Color online) Energy per neutron (MeV) as a function of density (fm$^{-3}$) 
of neutron matter for BSk17 and for the calculations of Ref.~\cite{fp81} (FP) and 
A18+$\delta v$+UIX$^*$ of Akmal et al.~\cite{apr98}.  }
\label{fig_nm}
\end{figure}

As seen in Fig.~\ref{fig_nm}, the energy-density curve of neutron matter for 
force BSk17 is identical to 
the realistic curve of Ref.~\cite{fp81} up to the supernuclear density of
0.4 fm$^{-3}$, as is the case with all our other forces that have been
fitted to $J$ = 30 MeV. It is to be noted that unlike Ref.~\cite{les06} we have
not had to resort to a second $t_3$ term in the Skyrme force in order to
simultaneously fit neutron matter and obtain the correct sign for the
isovector splitting of the effective mass.
Fig.~\ref{fig_nm} also shows the energy-density curve given by the realistic
calculation A18+$\delta v$+UIX$^*$ of Akmal et al.~\cite{apr98}. There is
more physics in this calculation than in the one of FP, but we are unable to
fit this curve without degrading the quality of the mass fit. However, there 
have been some very recent indications that this curve might be too steep 
\cite{li09}. 

\begin{figure}
\includegraphics[scale=0.3]{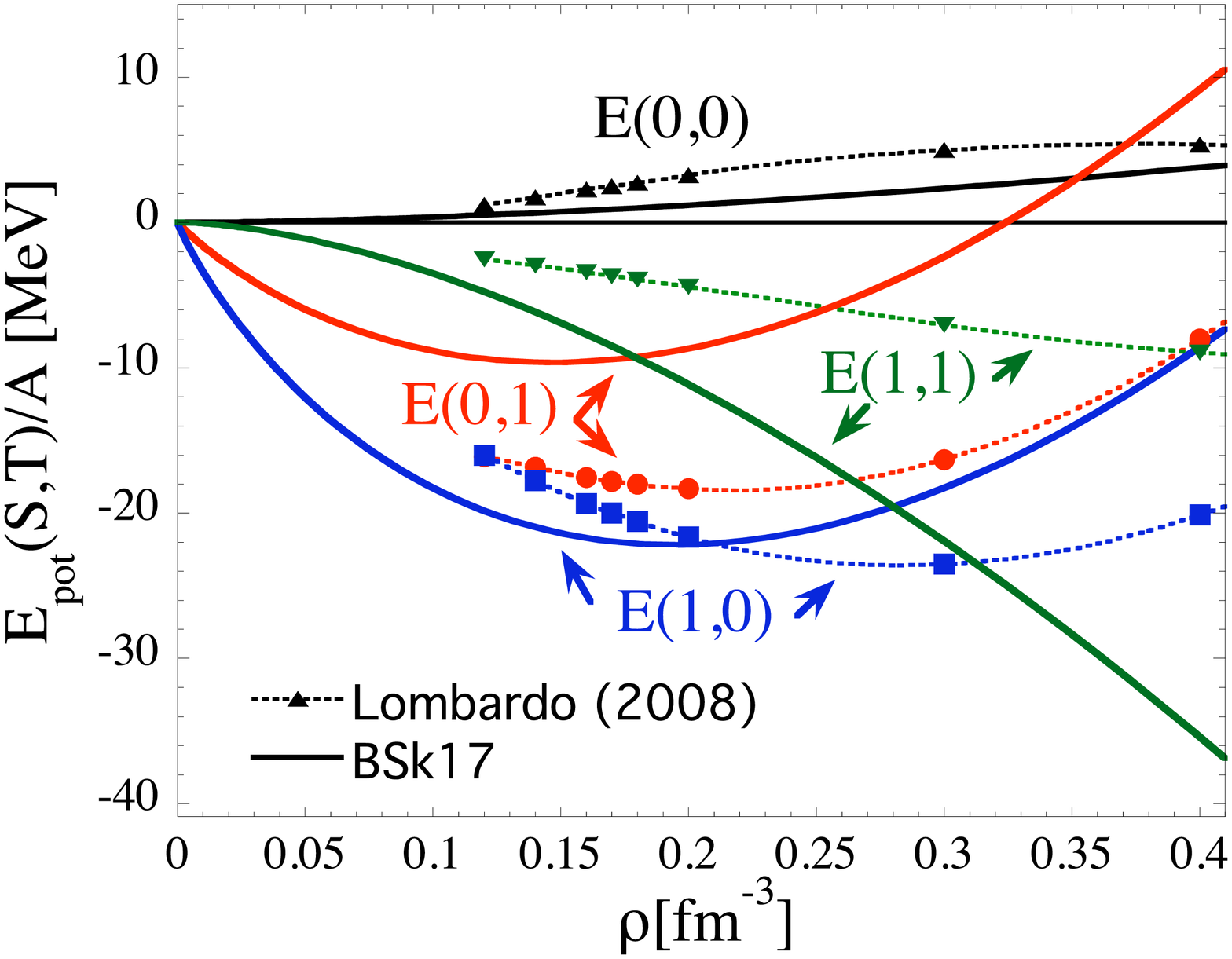}
\vskip -0.5cm
\caption{(Color online) Potential energy per particle in each $(S,T)$ channel as a function of
density for symmetric infinite nuclear matter. The full symbols (connected with the dashed lines) correspond to BHF calculations \cite{lom08} and  the solid lines to the BSk17 force.}
\label{fig_est}
\end{figure}

Fig.~\ref{fig_est} shows the potential energy per particle in each of the four
two-body spin-isospin $(S,T)$ channels as a function of density for symmetric
nuclear matter; we give results for both BSk17 and Brueckner-Hartree-Fock (BHF) 
calculations with realistic two- and three-nucleon forces \cite{lom08}. 
A fair agreement between BSk17 and the realistic calculations in all states can
be seen; note particularly that the deviation in the (1,1) channel is much less
marked than in Ref.~\cite{cha08}, mainly because there we compared with older
BHF calculations.

{\it Conclusions.} We have described
a new Skyrme-HFB nuclear-mass model, HFB-17, in which the 
contact pairing force is constructed from microscopic pairing gaps of
symmetric nuclear matter and neutron matter calculated from realistic
two- and three-body forces, with medium-polarization effects included. 
In this way the rms deviation with respect to essentially all the available 
mass data has been reduced, for the first time with a mean-field model, below
0.6 MeV. Given also the constraint imposed on the Skyrme force by
microscopic calculations of neutron matter, this new model is particularly 
well adapted to astrophysical applications involving a neutron-rich 
environment,
such as the elucidation of the r-process of nucleosynthesis, and the
description of supernova cores and neutron-star crusts.

\begin{acknowledgments}
S.G. and N.C. acknowledge financial support from FNRS.
\end{acknowledgments}

\end{document}